\documentclass[12pt,preprint]{aastex}
\usepackage{emulateapj5}
\usepackage{apjfonts}
\usepackage{epsfig}

\journalinfo{The Astrophysical Journal Letter}
\slugcomment{Accepted by The Astrophysical Journal Letters}

\def\calc{{\cal C}}

\def\esn{E_{\rm SN}}

\def\kms{\ifmmode {\rm km~ s^{-1}} \else {\rm km~s^{-1}}\ \fi}
\def\mbh{M_{\bullet}}

\def\mgii{\ifmmode Mg {\sc ii} \else Mg {\sc ii}\ \fi}
\def\oiii{\ifmmode [O {\sc iii}] \else [O {\sc iii}]\ \fi}
\def\feii{\ifmmode Fe {\sc ii} \else Fe {\sc ii}\ \fi}

\def\sigmagas{\Sigma_{\rm gas}}
\def\sigmastar{\dot{\Sigma}_*}
\def\sunm{M_{\odot}}

\def\vtur{V_{\rm tur}}

\def\gsim{\mathrel{\rlap{\lower 4pt \hbox{\hskip 1pt $\sim$}}\raise 1pt
\hbox {$>$}}}
\def\lsim{\mathrel{\rlap{\lower 4pt \hbox{\hskip 1pt $\sim$}}\raise 1pt
\hbox {$<$}}}
\def\lax{{$\mathrel{\hbox{\rlap{\hbox{\lower4pt\hbox{$\sim$}}}\hbox{$<$}}}$}}
\def\gax{{$\mathrel{\hbox{\rlap{\hbox{\lower4pt\hbox{$\sim$}}}\hbox{$>$}}}$}}

\begin{document}

\title{The starburst-AGN connection: the role of young stellar populations 
in fueling supermassive black holes}

\author{
Yan-Mei Chen\altaffilmark{1},
Jian-Min Wang\altaffilmark{1,2},
Chang-Shuo Yan\altaffilmark{1}, 
Chen Hu\altaffilmark{1}, and
Shu Zhang\altaffilmark{1}}

\altaffiltext{1}
{Key Laboratory for Particle Astrophysics, Institute of High Energy Physics,
Chinese Academy of Sciences, 19B Yuquan Road, Beijing 100049, China}

\altaffiltext{2}
{Theoretical Physics Center for Science Facilities, Chinese Academy of Sciences, Beijing 100049, China}


\begin{abstract}
  Tracing the star formation history in circumnuclear regions (CNRs)
  is a key step towards understanding the starburst-AGN
  connection. However, bright nuclei outshining the entire host galaxy
  prevent the analysis of the stellar populations of CNRs around
  type-I AGNs. Obscuration of the nuclei by the central torus provides
  an unique opportunity to study the stellar populations of AGN host
  galaxies. We assemble a sample of 10,\,848 type-II AGNs with a
  redshift range of $0.03\le z\le 0.08$ from the Sloan Digital Sky
  Survey's Data Release 4, and measure the mean specific star
  formation rates (SSFRs) over the past 100Myr in the central
  $\sim1-2$\,kpc . We find a tight correlation between the Eddington ratio
  ($\lambda$) of the central black hole (BH) and the mean SSFR,
  strongly implying that supernova explosions (SNexp) play a role in
  the transportation of gas to galactic centers.  We outline a model
  for this connection by accounting for the role of SNexp in the
  dynamics of CNRs.  In our model, the viscosity of turbulence excited
  by SNexp is enhanced, and thus angular momentum can be efficiently
  transported, driving inflows towards galactic centers. Our model
  explains the observed relation $\lambda \propto \rm
  SSFR^{1.5-2.0}$, suggesting that AGN are triggered by SNexp in CNRs.
\end{abstract}

\section{Introduction}
Much attention has been given to understanding the relation between
starbursts and AGN. Powerful emission from the central regions of
galaxies is generally thought to originate from gravitational energy
released via accretion onto supermassive BHs (Rees 1984), but, how to
fuel BHs from $\sim$kpc to pc scale remains an open question (Shlosman
et al. 1990; Wada 2004). On the other hand, the presence of starbursts
in CNRs has been conclusively established throughout the AGN family:
quasars (Brotherton et al. 1999; Hao et al. 2005, 2008), radio galaxies 
(Wills et al. 2002), Seyfert galaxies (Heckman et al. 1997; Le Floch et 
al. 2001; Gu et al. 2001; Imanishi 2002; Imanishi et al. 2003; Davies 
et al. 2007; Wang et al. 2007; Watabe et al. 2008) and low luminosity 
AGN (Cid Fernandes et al. 2004; Gonzalez-Delgado et al. 2004). 
Starbursts have even been considered the energy source of AGN
(Terlevichi \& Melnick 1985).  After intensive studies in the last 3
decades, especially with increasing evidence for the coevolution of
supermassive BHs and their host galaxies (Magorrian et al. 1998;
Tremaine et al. 2002), the debate of which process powers AGN emission
has become a question of the connection between starbursts and AGN
(e.g. the review by Heckman 2008).  The coexistence of these two
phenomena could simply reflect the fact that both live on the same
gas-based diet, however, the nature of the starburst-AGN connection
remains one of the mysteries in the evolution of galaxies.

Examining the starburst-AGN connection requires simultaneous
observations of nuclei and starburst regions. Thanks to the torus
obscuration, type-II AGN are the best laboratory for us to study the
connection of central BHs and their host galaxies (Cid Fernandes et
al. 2001).  
Statistical studies of large samples of galaxies in the SDSS reveal
systematic trends between host properties and BH activity (Cid
Fernandes et al. 2001; Kauffmann et al. 2003c, 2007; Wild et
al. 2007). The higher the luminosities/Eddington ratios of AGN, the
younger the ages of the stellar populations in the host bulges.  All
evidence from studies of type-II AGN undoubtedly indicates an
intrinsic connection between star formation and AGN. However, one key
puzzle remains unsolved: how to transport the gas over many
orders-of-magnitude in radius to feed the BH?

In this Letter, we analyse the star formation histories in the CNRs of
the host galaxies of type-II AGN, finding a strong correlation between
the Eddington ratios and the mean SSFRs over the past 100Myr. The
correlation explicitly implicates SNexp in driving the inflow
to BHs, providing a new clue to understand the processes occurring in
the CNRs. A model is proposed to explain the results with inclusion of
turbulence excited by SNexp. The cosmological parameters $H_0=70{\rm
  km s^{-1} Mpc^{-1}}$, $\Omega_{\rm M}=0.3$ and
$\Omega_{\Lambda}=0.7$ are used throughout the paper.

\section{Observational Connection: SSFRs vs Eddington ratios}
\subsection{The sample}
An available sample of type-II AGN from 
the SDSS-Data Release 4 (Adelman-McCarthy et al. 2004) 
can be found from the MPA/JHU catalog\footnote{The MPA/JHU catalog can be 
downloaded from http://www.mpa-garching.mpg.de/SDSS.} (Kauffmann et al. 2003c). 
This catalog includes those objects with H$\alpha$, [N {\sc ii}], 
\oiii and H$\beta$ emission lines detected with S/N$>$3. 
The objects were selected to be type-II AGN according to their line ratios: 
$\log ($[O {\sc iii}]/H$\beta) > 
0.61/\{\log ([$N {\sc ii}]/H$\alpha)-0.05\} +1.3$
or $\log ([$N {\sc ii}]/H$\alpha)>0.2$. Other
measurements of these type-II AGN and their host galaxies required in
this paper: dust extinction corrected \oiii$\lambda 5007$ luminosity ($L_{\rm
  [O~III]}$), stellar velocity dispersion ($\sigma$) of the bulge and
dust-corrected stellar masses ($M_*$), are also included in the
catalog. The stellar masses are estimated from D$_n$(4000) and
H$\delta_A$ by Kauffmann et al. (2003a) and the $3^{\prime\prime}$
fiber based measurements are scaled to the total galaxy luminosity.
We limit our sample to objects in the redshift range $0.03\le z\le
0.08$ and with $\log M_*/M_\odot \ge 10.5$, generating a sample of
10,\,848 type-II AGN. The redshift limits reflect our wish to study
the regions within bulges, taking advantage of the $3^{\prime\prime}$
aperture of SDSS fibers. The mass limit is set to avoid the
inclusion of disk dominated galaxies (Kauffmann et al. 2003c). The
median relative errors of $L_{\rm [O~III]}$, $\sigma$ and $M_*$ are
$12\%$, $6\%$ and $7\%$ respectively.

The rich stellar absorption line spectra of these type-II AGN provide
detailed information about their stellar content and
dynamics. However, they make the nebular emission-line measurements
difficult. Tremonti et al. (2004) designed a special-purpose code to
fit the stellar absorption-lines and continua. This code parameterises
the stellar populations of the host galaxies by fitting population
synthesis models, and makes the emission line measurements more
robust. We summarize the spirit of this code, then apply the results
to study the stellar populations of type-II AGN hosts in \S2.2.

In order to quantify the connection between the star formation and BH
activity, we must estimate the BH masses and accretion rate of the
galaxies. The BH masses can be estimated from stellar velocity
dispersion by
$\log\left(\mbh/\sunm\right)=8.2+4.02\log\left(\sigma/200{\rm
    km~s^{-1}}\right)$ (Tremaine et al. 2002). Following the method
described in Heckman et al. (2004), we correct $L_{\rm [O~III]}$ in
each AGN for the contribution from star formation. Then, given the
bolometric luminosity $L_{\rm Bol}\approx 3500 L_{\rm [O~III]}$
(Heckman et al. 2004), the Eddington ratio can be obtained by
\begin{equation}
\lambda=\frac{L_{\rm Bol}}{L_{\rm Edd}}\approx 0.25~L_{41}M_7^{-1},
\end{equation}
where $L_{\rm Edd}=1.38\times 10^{38}\left(\mbh/\sunm\right){\rm ergs~s^{-1}}$ is the 
Eddington luminosity, $M_7=\mbh/10^7\sunm$ and $L_{41}=L_{\rm [O~III]}/10^{41}{\rm ergs~s^{-1}}$.
We find that 90\% of BHs have masses larger than $10^7\sunm$, with
most between $10^7\sim 10^8\sunm$. The \oiii luminosity distribution
spans from $3\times 10^{39}$ to $3\times 10^{42}$ergs~s$^{-1}$,
implying a range of bolometric luminosity from $10^{43}$ to $10^{46}$
ergs~s$^{-1}$. The Eddington ratios are homogeneously distributed from
$10^{-3}$ to 1, about $90\%$ objects are in the range $10^{-2}$ to 0.3,
indicating that the current sample covers a relatively complete range
of the standard accretion disks (Shakura \& Sunyaev 1973). In one
word, the current sample represents an homogeneous one in the relevant
parameters.

It should be mentioned that aperture effects are negligible in this
work, we have verified that a sample limited to a narrower redshift
range of $0.03\sim 0.05$ produces almost the same results. The
$3^{\prime\prime}$ SDSS fiber aperture corresponds to a physical
radius of $\sim 1-2$kpc for the galaxies in the present sample,
allowing us to focus on the physical processes in the CNRs.

\subsection{Stellar population synthesis}
The continua and absorption lines of each galaxy are fitted by a
stellar population model (Tremonti et al. 2004), under the basic
assumption that any galaxy star formation history can be approximated
as a sum of discrete bursts. The library of template spectra is
composed of single stellar population models generated using the
population synthesis code of BC03 (Bruzual \& Charlot 2003), including
models of ten different ages (0.005, 0.025, 0.1, 0.2, 0.6, 0.9, 1.4,
2.5, 5, 10Gyrs) and three metallicities (0.2$Z_{\sun}$, $Z_{\sun}$,
and 2.5$Z_{\sun}$). The template spectra are convolved to the measured
stellar velocity dispersion of each SDSS galaxy, and three best
fitting model spectra with fixed metallicity are constructed from a
non-negative linear combination of the template spectra, with dust
attenuation modeled as an additional free parameter.  The metallicity
which yields the minimum $\chi^2$ is selected for the final best-fit
model. The median $\chi^2$ is 1.2 in our sample. The fitting results
can be found on the SDSS-MPA webpage.

We split the sample into two bins by galaxy mass at
$10^{10.8}\sunm$. The bins have roughly equal numbers of objects
and median stellar masses of $10^{10.6}\sunm$ and $10^{11.0}\sunm$.

In Fig. 1, we plot the fraction of different stellar populations for
these two mass bins.  The X-axis represents ten ages of the model
stellar populations as indicated in the caption. From left to right,
age increases. The coloured lines indicate different Eddington
ratios. The Poisson errors on the histogram bins are negligible. We
see that, at a given $\lambda$, the distribution of the fraction of
stellar populations is roughly the same for the two mass bins.  The
fraction of young stellar populations ($\le 100$Myr) increases with
Eddington ratio, whereas the fraction of the old populations ($\ge
10$Gyr) decreases. The stellar populations are more dependent on the
Eddington ratios than galaxy masses. At a certain $\lambda-$bin, we 
test the similarity of the distribution for the two mass bins.  
The two-sided K-S test gives a probability value greater than 
97\%  for all the four $\lambda-$bins. 

The SDSS spectra are obtained with optical fibers, therefore we
can only measure the spatially-integrated stellar populations. To
our knowledge, any evidence for radial gradients of stellar
populations in CNRs remains controversial. In Seyfert galaxies, the
CNR ring-shaped starbursts (e.g. Storchi-Bergmann et al. 1996) are
found to be more powerful than the nuclear ($\sim 100$pc) starbursts
(Le Floch et al. 2001; Imanishi 2002). On the other hand, Raimann et
al. (2003) use long-slit optical/near UV spectroscopy to show that
there is no strong systematic radial gradient of the stellar
populations on scales from nuclear regions to a few kpc for a sample
of 22 type-II Seyfert galaxies. It is fortunate, however, the SDSS
data do allow us to examine the primary driver of the starburst-AGN
connection.

\vglue 0.3cm
\figurenum{1}
\centerline{\psfig{figure=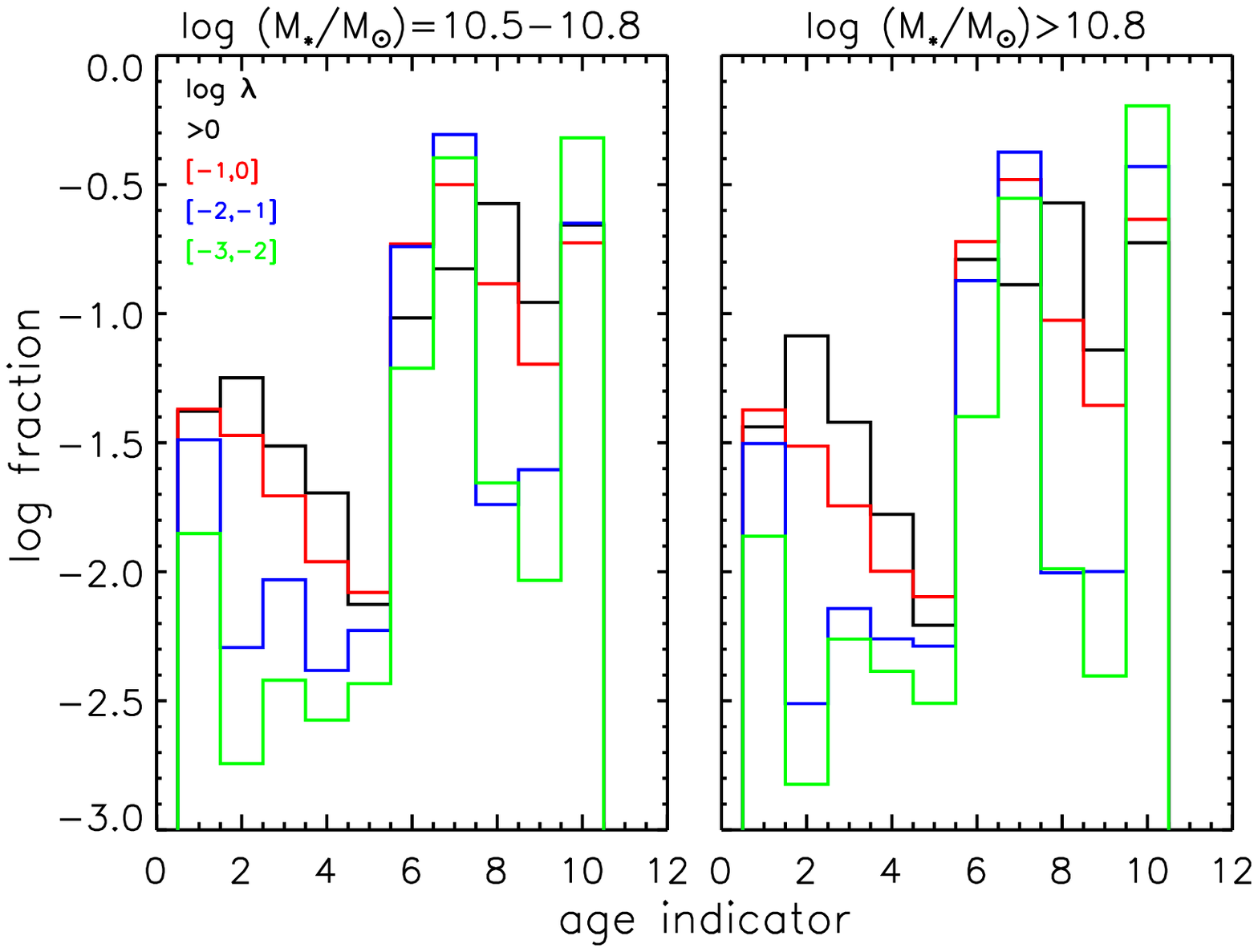,angle=0,width=8.5cm}}
\figcaption{\footnotesize The histogram distribution of the fraction
  of stellar populations in type-II AGN. The X-axis represents ten
  ages (0.005, 0.025, 0.1, 0.2, 0.6, 0.9, 1.4, 2.5, 5, 10Gyrs) of the
  model stellar populations from left to right.  The black, red, blue
  and green lines correspond to samples with different Eddington
  ratios.  It is found that black holes with higher Eddington
  ratios have younger stellar populations. }
\label{fig1}
\vglue 0.5cm

\subsection{Stellar populations and BH activities}
The mean specific star formation rate over the past 100Myr can be
estimated from fitting the stellar continua of the host galaxies
\begin{equation}
{\rm SSFR(<0.1Gyr)}=\frac{\rm SFR}{M_*}=\frac{\sum_{i=1}^{3}f_i}{0.1{\rm Gyr}},
\end{equation}
where $f_i$ is the fraction of stars in each age bin and the sum is
over the first three age bins i.e. over the past 100Myrs. We note that
stars with masses $\sim 6.0\sunm$ have an main-sequence lifetime of
100 Myrs.  Fig. 2 shows the relation between the mean SSFRs and the
Eddington ratios of the central BHs.  We find that galaxies with
higher specific star formation rates in the CNRs have higher Eddington
ratios of the central BHs. A linear regression shows
\begin{equation}
\log\lambda = (-0.73\pm 0.01)+(1.50\pm 0.01)\log {\rm ~SSFR},
\end{equation}
with the Pearson's correlation coefficient of 0.53 and a probability
of $<10^{-5}$ that this could be obtained by chance.  This strong
correlation is striking in that: 1) it directly quantifies the physical
starburst-AGN connection; 2) the relation significantly deviates from
the linear relation ($\lambda\propto {\rm SSFR}$); 3) it strengthens
the case for the role of starbursts in activating BHs, and the
potential role of SNexp in particular.

This correlation provides strong constraints on physical models. The
accreted mass onto BHs is only a tiny fraction ($\sim 10^{-3}$) of the
total gas, a simple relation like $\lambda\propto {\rm SSFR}^q$ is
then expected if starbursts and AGN simply have the same diet, where
$q$ is unknown. A non-linear index $q \neq 1$ implies a more
complicated interaction between the star formation and AGN. A higher
SSFR simply means higher rates of SNexp.  The present high$-q$ value
($q>1$) might stress a potential role of SNexp in triggering AGN.  As
a source of energy in the CNRs, SNexp are playing, at least, two
roles: 1) heating the medium; 2) exciting turbulence via dissipation
of their kinetic energy. The first may in principle lead to a certain
level of star formation suppression, whereas the second role is that
the excited turbulence efficiently transports angular momentum and
drives inflows. AGN are then triggered. In the next section, we show
how detailed
processes occurring in these regions can be examined by the relation
between $\lambda$ and SSFR in a quantitative way.

\vglue 0.5cm 
\figurenum{2}
\centerline{\psfig{figure=fig2.ps,angle=0,width=8.5cm}}
\figcaption{\footnotesize Relation between the specific star formation
  rates and the Eddington ratios.  A simple regression, the blue line,
  gives $\lambda \propto {\rm SSFR}^{1.5}$, which is consistent with
  the expected relation of our theoretical model (eq. 9). The red curves
  are density contours. The green dashed line which passes through the
  contours gives $\lambda \propto {\rm SSFR}^{2.0}$. The magenta and
  cyan histograms are for mass bins of $>10^{10.8}\sunm$ and $\le
  10^{10.8}\sunm$, respectively. The typical error bars are plotted in the
  upper-left corner, and are estimated from uncertainties in $L_{\rm
    [O~III]}$, $\sigma$ and stellar population fitting.}
\label{fig2}
\vglue 0.5cm

\section{Comparison with theoretical models}
As we argue in \S2, the correlation shown in Fig 2. clearly unveils an intrinsic connection between 
the star formation and  triggering of AGN. Currently, there are two main kinds of theoretical models for 
this connection, represented by Thompson et al. (2005, hereafter T05) and Kawakatu \& Wada (2008, 
hereafter KW08). T05 show that the radiation pressure on dust grains is 
able to support the star forming regions, but in turn the star formation is 
controlled by a self-adjustment of the Toomre$-Q$ parameter. This model assumes that
the gaseous medium is homogeneous and stresses a balance between 
supernova heating and self-gravity in the star forming region.  Unfortunately, this model makes the 
temperature of the gas so low that the commonly used $\alpha-$viscosity is not strong enough to 
transport angular 
momentum since the sound speed is too slow. An external torque is needed to transport angular 
momentum to fuel BHs, and bar within bar instability is proposed. However, this suggestion 
lacks observational support. In a statistical sense, at least in the modern universe 
($z<1$), there is no evidence for the connection between star formation/AGN activity and 
lopsidedness (Li et al. 2008; Reichard et al. 2008).  

High resolution hydrodynamical simulations show the ISM gas
in the central 2 Kpc is multi-phase and not steady and homogeneous,
and that the global geometry is supported by internal turbulence
likely caused by SNexp (Wada \& Norman 1999, 2001, 2002; Korpi et
al. 1999 for $\sim $100 pc).  The vertical structure is supported by
turbulence (see also Vollmer \& Beckert 2003; Collin \&Zahn
2008). KW08 propose a geometrically thick, clumpy and
turbulence-dominated disk supported by energy from SNexp.  Mass rates
of accretion onto BHs may, in principle, strongly depend on star
formation rates.  According to eq. (12) in KW08, $\lambda \propto {\rm
  SSFR}$ is expected, which deviates from the observed relation. But
the KW08 model addresses the intrinsic connection between starbursts
and AGN, namely the role of SNexp in the transportation of angular
momentum.

We assume that turbulence developed from SNexp is responsible for transporting angular momentum and 
is energized by
\begin{equation}
\sigmagas \vtur^2=\eta \sigmastar \esn \left(\frac{H}{V_{\rm tur}}\right),
\end{equation}
where $\sigmagas$ is the gas surface density, $V_{\rm tur}$ the
turbulence velocity, $\sigmastar$ the star formation rate surface
density, $E_{\rm SN}$ the energy per SNexp, and $H$ is the height of the gas disk.
Here $\eta$ is the conversion efficiency of the explosion 
energy into turbulence and highly uncertain. The accretion rate onto BHs is then given by
\begin{equation}
\dot{M}_{\bullet}=2\pi \nu \sigmagas \Omega^{\prime},
\end{equation}
where $\Omega^{\prime}=d\ln \Omega(R)/d\ln R$, $\Omega=\left(GM/R^3\right)^{1/2}$, 
$M$ is the total mass within radius $R$, and $\nu$ is the kinetic viscosity. We also assume that the 
turbulence causes the viscosity, namely,
\begin{equation}
\nu=\alpha \vtur H,
\end{equation}
where $\alpha$ is a constant. Similar to Vollmer \& Beckert (2003) and Collin \& Zahn (2008), 
we introduce the Toomre$-Q$ as a free 
parameter to describe the gravitational instability
\begin{equation}
Q=\frac{\Omega^2}{\sqrt{2}\pi G\calc\rho_{\rm gas}}
 =\frac{\sqrt{2}\Omega^2H}{\pi G\calc\sigmagas},
\end{equation}
where $\calc$ is the clumpiness of gas and $\rho_{\rm gas}=\sigmagas/2H$ is used. We then have
\begin{equation}
\dot{M}_{\bullet}=c_0
		  \alpha Q^{4/3}\Omega^{-8/3}\sigmastar^{(6+\beta)/3\beta},
\end{equation}
where the Kennicutt-Schmidt law $\sigmastar=\Sigma_0\sigmagas^{\beta}$ is assumed and the constant
parameter
$c_0=2^{1/3}\pi^{7/3}\left(\eta\esn\right)^{1/3}\left(G\calc\right)^{4/3}\Sigma_0^{-2/\beta}\Omega^{\prime}$.
We find that the accretion is 
strongly correlated with the starburst rates as $\dot{M}_{\bullet}\propto \sigmastar^{1.8}$ 
for $\beta=1.4$ (Kennicutt 1998). This shows a clear relation between
the star formation and BH activity.

Employing the $M_*-R$ relation of $R\propto M_*^{\gamma}$, we have 
$\sigmastar\propto {\rm SFR}/R^2= {\rm SSFR}~M_*/R^2= {\rm SSFR}~M_*^{1-2\gamma}$,
where we assume $M_*=f_*M$ and $f_*$ is a constant. With the help of the Magorrian relation of 
$\mbh\propto M_*$, we have $\lambda=L_{\rm Bol}/L_{\rm Edd}\propto \dot{M}_{\bullet}/M_{\bullet}$
\begin{equation}
\lambda
       \propto M_*^{0.48\gamma-0.57}Q^{4/3}{\rm SSFR}^{1.8}
       = M_*^{-0.3}Q^{4/3}{\rm SSFR}^{1.8},
\end{equation}
where the index is about $\gamma\approx 0.55$ (Shen et al. 2003). This
relation ($\lambda\propto {\rm SSFR}^{1.8}$) agrees with the results
in Fig. 2, which gives a power index in the range of $1.5\sim 2.0$.
The weak dependence of $\lambda$ on $M_*$ also follows from
eq. (9) in agreement with the histogram distributions of $\lambda$ for
the two mass bins (the higher $M_*$, the lower $\lambda$ from the
histogram plot). We note that the strong correlation has significant
scatter. We would ascribe this scatter to the free parameter $Q$,
which numerical simulations have shown can vary widely (Wada \&
Norman 1999, 2002; Vollmer \& Beckert 2003). It could also 
be due to a slight mis-match in timescales of our SSFR measure 
and AGN triggering. We would like to stress
here that the agreement of eq. (9) with the correlation (3) results
from turbulent viscosity excited by the SNexp.

The strong starburst-AGN connection provides new evidence for the
coevolution of black holes and galaxies. We need a set of time-dependent 
equations for the interaction between AGN and starbursts in clumpy 
CNRs, as well as the duty cycle of quasars (Wang et al. 2006; 2008), 
taking into account the role of SNexp.  The time-dependent model 
must also address the transition
from blue to red galaxies driven by AGN, and incorporate a
connection with downsizing evolution of galaxies (e.g. Chen et
al. 2009). Observationally, more detailed information of radial
gradients of stellar populations from long-slit or
integral-field-unit spectrographs is expected to provide more
constraints on theoretical models.

\section{Conclusions}
We show a striking correlation between the Eddington ratios and
specific star formation rates in a large sample of type-II AGN. We
find that type-II AGN in general show young stellar populations, and
that the higher the fraction of young stars, the higher the Eddington
ratio. This intrinsic connection strengthens the evidence for the
role of SNexp in triggering BH activities. The measured correlation is
consistent with a scenario in which turbulence excited by SNexp is
responsible for the transportation of gas to the central AGN.

Future studies on the starburst-AGN connection are needed using high
quality data to study the correlations between $\lambda$ and SSFR for
different classes of galaxies in terms of their morphologies and
colors. This will give stronger constraints on the details of AGN triggering.

\acknowledgements{The authors are grateful to the referee for the helpful report improving the manuscript. 
V. Wild is greatly thanked for many useful suggestions and help in this work.
We acknowledge interesting discussions with S. N. Zhang. We appreciate the stimulating 
discussions among the members of IHEP AGN group. The research is supported by NSFC and CAS via NSFC-10733010 
and 10521001, KJCX2-YW-T03, and the National Basic Research Program of China (2009CB824800), respectively.}

\end{document}